# Femtosecond-laser-induced nanoscale blisters in polyimide thin films through nonlinear absorption

Alan T. K. Godfrey*, L. N. Deepak Kallepalli, Jesse J. Ratté, Chunmei Zhang* and P. B. Corkum

Joint Attosecond Science Laboratory, University of Ottawa and the National Research Council, 25 Templeton St., Ottawa, K1N 6N5, Canada

*Corresponding Emails: agodfrey@uottawa.ca, chunmei.zhang@uottawa.ca

**Abstract:**

Nonlinear absorption of femtosecond laser pulses provides a unique opportunity to confine energy deposition in any medium to a region that is below the focal diameter of a pulse. Illumination of a polymer film through a transparent high bandgap material such as glass, followed by nonlinear absorption of 800 nm light in polymers, allows us to further restrict absorption to a very thin layer along the propagation direction. We demonstrate this confinement by simulating femtosecond-laser-induced polymer modification by linear, two-photon and three-photon absorption, and discuss the control over energy absorption in polymers that multiphoton processes offer. Energy deposited behind a thin polymer film induces a protruding blister. We present experimental results of blister diameter and height scaling with pulse energy. Using 0.95 NA focussing, we obtained laser-induced blisters with diameters as small as 700 nm suggesting blister-based Laser-Induced Forward Transfer is possible on and below the single-micron scale. Sub-micrometer blister formation using femtosecond lasers also offers a novel method of direct, precise laser-writing of microstructures on films with single laser pulses. This method is a possible alternative to lithography, laser milling, and laser-based additive machining which leaves the surface composition unchanged.

**Introduction:**

When a femtosecond light pulse irradiates a material in vacuum, there is no limit to the intensity incident on the vacuum-material interface. One can easily exceed the ablation threshold of the material, leading to material removal long before heat transport becomes important. This allows, for example, deterministic sub-focal-spot machining or nanoscale fabrication.

A similar situation can arise for irradiation of a low-bandgap material by light that has passed through a high-bandgap medium. For example, a femtosecond pulse containing ~1.5 x $10^{13}$ W/cm$^2$ can pass through a thick fused silica plate before free carrier generation in the medium is great enough to significantly attenuate the beam. A few-cycle pulse can even reach ~$10^{14}$ W/cm$^2$ [1]. If high intensities are reached while staying below the critical power for self-focussing (by using tight focussing or a sufficiently-thin medium), a polymer film on silica can experience high-order multiphoton absorption of a well-controlled beam. Furthermore, with a modest increase of intensity above 1.5 x $10^{13}$ W/cm$^2$, the primary influence of free-carrier generation is to cap the intensity but leave the pulse otherwise unchanged [2].

We study the light-polymer interaction under these conditions. When an intense laser pulse is focussed through a substrate onto a coated film, it can create a pocket of superheated material beneath the surface of a film that expands into a protruding blister Blisters have been used to achieve Laser-Induced Forward Transfer, where they impart thrust on an object or material on the surface, thereby desorbing it while avoiding direct laser exposure [3].

Both polymers and metals have been explored as sacrificial laser-absorbing layers with pulse durations ranging from nanoseconds to femtoseconds [3]–[8]. Polymer films are ideal for preserving the chemical purity of the transferred material; metals are prone to degradation from thermal and chemical damage, leading to contamination during transfer [9]. Arnold et al. have demonstrated blister formation by linear absorption of nanosecond lasers in polyimide films [3], [5], [8]. The underlying physics of polymer blister formation in the nanosecond pulse regime was addressed for blisters on the scale of 10 μm to 100 μm in width. However, no thorough studies regarding femtosecond lasers and nonlinear absorption processes have been performed to our knowledge. Blisters on the few-micron and sub-micron scale have also not been explored. Nonlinear absorption of femtosecond pulses leads to confined energy deposition, due to threshold-like absorption behaviour and lack of heat dissipation over the timescale of the pulse.

In this paper, we demonstrate the advantages of nonlinear absorption of femtosecond pulses to create polymer blisters. We calculate material breakdown induced by single femtosecond pulses in experiment-like conditions to illustrate the confinement of energy deposition. We then show blisters fabricated in polyimide films using single femtosecond pulses, which are the first made at the sub-wavelength scale. We examine the effects of laser pulse energy on the resulting blister size, and show a linear relationship between the energy deposited into the polymer and the resulting blister volume.

This work establishes a novel method for direct and precise laser-writing of microstructures on films using single laser pulses, which presents an alternative to lithography, laser milling, and laser-based additive machining. Since the laser energy is deposited beneath the film, morphological changes are achieved while preserving the surface chemistry [10].

**Experimental & Calculation Details:**

*I – Experimental*

The schematic of the femtosecond-laser-induced blister formation experiment is shown in **Fig. 1(a)**. We use a Ti:sapphire laser producing transform-limited pulses with durations of $\tau = 45$ fs (FWHM) at a central wavelength of 800 nm. The beam was passed through a spatial filter and the spatial profile was verified to be Gaussian using a beam profiler. We adjusted pulse energy using a rotatable half-wave plate followed by a linear polarizer. The laser was focussed by a 0.4 NA microscope objective, which was mounted into a vertical motor stage for adjusting focal spot placement. The 6 mm aperture of the objective was filled with a 4 mm diameter beam; therefore, the 1/e² focal spot diameter is estimated to be $d_0 = \left(\frac{6 \text{ mm}}{4 \text{ mm}}\right)\frac{1.22\lambda}{NA} \approx$ 3.7 µm [11], [12]. We adjusted the chirp using a grating compressor to maximize the intensity after the objective, indicated by second harmonic yield from a BBO crystal. In one instance, a 0.95 NA with a 10 mm entrance aperture was also used, resulting in an estimated focal spot diameter of 2.6 µm. Reduced intensity from additional chirp by the 0.95 NA objective was compensated by increasing the pulse energy.

We use polyimide films on #1.5 Fisherbrand borosilicate glass coverslips as substrates, which were pre-rinsed in acetone, isopropyl alcohol and deionized water and dried on a hotplate. We calculate the critical power of borosilicate glass at 800 nm to be 2.5 MW [13]; for our 50-fs pulse, this corresponds to a pulse energy of 180 nJ. Focussed to a 3 µm spot size, the light intensity before self-focussing in the substrate occurs can reach 3 x $10^{13}$ W/cm² in the absence of nonlinear absorption by the substrate. Films were spin-coated using PI-2555 precursor purchased from HD Microsystems, at a spin speed of 6000 RPM for 60 seconds. This yielded a film thickness of 1.3 µm, as determined by contact profilometry. We measured the linear absorption spectra of polyimide films from 280 to 800 nm, shown in **Fig.1(b)**. Films were transparent to wavelengths above 500 nm, indicating that 800 nm light can only be absorbed through two-photon absorption and higher-order processes. The absorption spectra of the polyimide film match the literature and the absorption is ascribed to an n → π* transition [14], [15].

Above the lens stage, samples are mounted onto a two-axis horizontal stage. Samples are oriented film-side-up, so that the laser is focussed through the coverslip onto the underside of the polyimide film. A dichroic mirror is used before the objective in a coaxial geometry, allowing a small fraction of laser light to reflect from the sample, re-collimate through the objective, and exit to an imaging line for in-situ laser spot monitoring. Coupling in white light and changing the position of the objective also allows for in-situ white-light microscopy. We used in-situ imaging to find the optimal position of the laser focus on the sample, by firing pulses with energies near the damage threshold while adjusting the objective's position.

The laser was used in single-shot mode to generate individual blister structures in polyimide films. We characterized the resulting blisters using a JPK atomic force microscope (AFM). From the AFM data, we determined the height, base diameter, and volume of each blister. Blister height is defined as the difference between local maximum and minimum height values, and diameter is defined as effective circular diameter of the base area for each blister. Blister volume is defined as the volume between the blister's surface and the plane of the unmodified film.

*II – Calculation*

To gain a qualitative understanding of the light-polymer interaction, we consider the energy absorption profile when a femtosecond pulse causes laser-induced breakdown through linear, two-photon, and three-photon absorption in a polymer film. Treating each order of absorption separately, we highlight their main features. We assume that the film has the physical characteristics of polyimide ($\rho = 1.42 \text{ g cm}^{-3}$, $c = 1.09 \text{ J g}^{-1} \text{ K}^{-1}$, $T_{decomp} = 550°C$) [16], [17]. We choose the linear absorption coefficient to match previous works on blister formation in polyimide through linear absorption of nanosecond pulses, for comparison with nonlinear absorption [3]. While the heat capacity will not be constant over the temperature range the polymer will experience, we will treat it as constant in our qualitative model. Since nonlinear absorption coefficients of undoped polymers are not reported, we chose two- and three-photon absorption cross sections to be those of zinc oxide [18] – a material with a similar band gap. We also assume that the laser pulse has a Gaussian temporal profile with $\tau = 50$ fs being the 1/e full width of the intensity in time, and spatial profile $2\omega_0 = 3$ µm where $2\omega_0$ is the 1/e full width of the electric field in space. Further details of this calculation are contained in **Appendix A**.

**Results and Discussion:**

In **Fig. 2**, we show the calculated temperature profile due to each of the three absorption mechanisms. Pulse energies were chosen such that the peak temperature in the material reaches 10,000K in all cases. This is the temperature of cold-dense plasmas [19] which is accessible, at least briefly, under normal lab conditions. At this temperature we expect molecular dissociation and ionization. Thus, within this assumption, we heat the polymer near the interface until it forms a plasma, transforming its chemical state in the process [10], [20]–[23].

**Figure 2** shows that the heat deposition region is smaller in all directions for nonlinear absorption. Most notably, the depth of energy deposition is more than 20 times shallower in the case of the three-photon absorption, compared to linear absorption, and approximately 2/3 of the diameter. The pulse energy used in this case was also an order of magnitude lower; this is owed to the high intensities of tightly-focused femtosecond pulses and the cubic intensity dependence of three-photon absorption. **Figure 2** also illustrates more general characteristics of energy deposition through high-order nonlinear absorption. After the third-order term dominates (for intensities above $10^{12}$ W/cm$^2$), most of the beam energy is deposited within a very small volume. The penetration depth is very shallow, thereby resulting in precise energy deposition within a thin film. Higher intensity, which lead to higher orders of absorption, will accentuate this trend without changing the overall conclusion. These characteristics are especially important for blister-based LIFT of sensitive materials on the nanoscale.

Now we move to experiment. We will show that, once we exceed an intensity of 3 x $10^{12}$ W/cm$^2$ in experiment (where three-photon absorption dominates in our model), we create blisters with volumes that grow linearly with intensity. For even higher input intensity, this growth slows since intensities above the ionization threshold of the substrate are now attenuated before reaching the polymer [2]. However, we will show that, by estimating the energy deposited in both the dielectric and the polymer, the linear growth of the blister volume with the energy absorbed by the polymer remains valid.

We first examine the dependence of blister height and diameter with pulse energy. **Figure 3** shows AFM scans of blisters created near the blister formation threshold using a 0.4 NA objective. Along the X-direction, four blisters were made at a fixed pulse energy to assess repeatability of the blister formation process. Along the Y-direction, pulse energy was varied. We observe the blister formation threshold to be approximately 19 nJ of pulse energy (peak fluence of 0.09 J/cm$^2$, peak intensity of 4 x 10$^{12}$) as measured after losses from the microscope objective. This pulse energy resulted in a blister of 75 nm height and 1.2 µm diameter. At low energies, blisters scale strongly in both height and diameter with pulse energy. For example, pulse energies three times higher than the threshold resulted in blisters of 1 µm height and 3.9 µm diameter.

**Figure 4** shows blisters formed in the same conditions but with further increases in pulse energy. In this regime, blisters have a weaker, linear-like scaling in both height and diameter, compared to the strong nonlinear scaling near the threshold. In the rightmost column of **Fig. 4**, pulse energies of approximately 11 times the threshold energy yield blisters with heights of 2 µm and base diameters of 5.5 µm. As pulse energy is increased to 225 nJ and above, blisters rupture due to excessive pressure built up beneath the film; they are no longer left intact, showing cracking, diminished height, and removal of material.

Height and diameter scaling for intact blisters in these experiments is summarized in **Fig. 5**. As also seen directly through AFM images, both scaling curves have two distinct regimes: {1} nonlinear scaling from the threshold pulse energy to 75 nJ, and {2} linear-like scaling onward until the onset of rupture. The monotonic increase of blister height and diameter with pulse energy is consistent with the literature, owing to increased temperatures and pressures of the material confined beneath the film. See **Appendix B** for further details and characterization of blister rupture.

Since the laser pulses in these experiments are often intense enough to cause substrate breakdown, we must account for energy lost to the substrate before reaching the polymer. In these situations, any intensity exceeding the breakdown threshold of the medium is simply removed from the beam to a reasonable approximation [2]. We adopt this model for a Gaussian pulse in both space and time. We estimate the intensity threshold for blister formation as the peak intensity in time at the edge of a blister formed using the threshold pulse energy ($I_{peak} \approx 3 \times 10^{12}$ W/cm$^2$). We then calculate the energy absorbed in the polymer as a function of the pulse energy used. This decouples the effects of nonlinear optical interactions in glass and polyimide from the expansion process of polyimide blisters.

**Figure 6** shows the dependence of blister volume, height, and diameter on energy absorbed in the film, accounting for absorption losses in the substrate. The resulting blister volumes ($V$) shown in **Fig. 6(a)** are linear with the amount of energy deposited in the film ($E$), to a good approximation. **Figure 6(b)** shows that height ($H$) and diameter ($D$) scale approximately with the square root and 4$^{th}$ root, respectively, of the absorbed pulse energy. Considering that a conical-like volume is proportional to base area ($\sim \frac{1}{4}\pi D^2$) times height, the sum of these powers is consistent with the linear trend seen for blister volume.

The linear trend between blister volume and energy absorbed in the film can be physically motivated. Since the laser-induced transition from solid to a dense plasma is well-established at intensities above $3 \times 10^{12}$ W/cm$^2$, we can estimate that most pulse energy is spent creating the plasma and transporting heat. This dense plasma contains many dissociated products, and will also thermally break down the neighbouring polymer. Through these processes, many new products are created, some of which are gaseous. As we add more energy, we break proportionally more bonds, which creates proportionally more gas. The gas we create will be at an elevated temperature and pressure but a very small volume. This is followed by an isothermal expansion process. Since we know the volume of gas created, it should be possible to estimate the time-dependent pressure and therefore, the surface motion, which is a measurable quantity.

A powerful consequence of nonlinear absorption is the possibility of energy deposition at scales below the diffraction limit. To understand the resolution advantage of forming blisters through nonlinear absorption, we closely examine AFM images of the smallest blisters formed using 19 nJ of pulse energy and a 0.4 NA objective (as shown in **Fig. 3**). The smallest of these is shown in **Fig. 7(a)**. The resulting blisters have $\frac{1}{e^2}$ diameters of $1.02 \pm 0.1$ µm, which is 28% of the focal spot diameter. The $\frac{1}{e^2}$ diameter definition is used because blister structures have smooth curvature and no distinct cut-off, like the intensity distribution of the Gaussian focal spot used to create them. We also examine the smallest blister made using 170 nJ of pulse energy and a 0.95 NA objective, shown in **Fig. 7(b)**. In this case, a diameter of 710 nm was achieved, which is again 28% of the focal spot diameter, and is smaller than the laser wavelength.

Looking forward, even smaller structures should be possible. A key consideration is the thickness of the irradiated film; the resolution limit in sub-wavelength desorption of thin films decreases significantly with decreasing film thickness [24]. We have calculated, and experimentally shown through helium ion microscopy and X-ray photoelectron spectroscopy, that the penetration depth into the film becomes very small for high-order nonlinear absorption, which will aid in creating intact blisters in thinner films [10]. Higher orders of nonlinear absorption could also be used to further restrict energy deposition.

**Conclusions:**

Femtosecond laser pulses provide a unique opportunity to deposit energy in a highly-controlled manner, through choice of absorption mechanism, pulse energy, pulse duration, and substrate mediation via intensity and material choice. Through our calculations, we have illustrated the advantage of femtosecond pulses for confining energy deposition in a thin film. We have demonstrated this advantage by achieving the first laser-induced blisters smaller than the laser wavelength. Blister formation at these scales is highly tunable, with resulting blister volumes that scale linearly with energy deposited in the material.

Further steps could be taken to refine the blister formation process. For example, a composite film with stacked layers of different materials could be used to control energy deposition and expansion separately, and we could employ high-speed surface interferometry to measure the height and velocity of a blister as it expands. In addition, energy deposition will scale linearly with pulse duration for a fixed peak intensity, providing further control over expansion rate. These new techniques would be a novel and convenient way to tailor blister-based Laser-Induced Forward Transfer of sensitive materials, where ejection speed is critical.

Precise fabrication of laser-induced blisters could impact several fields. This approach could lead to nanoscale blister-based Laser-Induced Forward Transfer of arbitrarily sensitive materials. Blisters on the few-micron and sub-micron scale can also be used for direct surface texturing and patterning of materials. Precise micro-texturing and perturbation of surfaces can be achieved without lithography, laser milling, or laser-based additive machining. No external processes are used to add or remove materials, and the surface composition remains unchanged since the energy is confined beneath the film. Blister microstructuring of surfaces could find interesting applications in chemically-stable superhydrophobic surface patterning and rapid microlens array fabrication. Blisters may also be used to create metamaterials by directly writing sub-wavelength surface structures.

**Funding:**

We acknowledge funding from the following sources: Natural Sciences and Engineering Research Council of Canada (NSERC) Engage (EGP 523138-18); and Discovery(RGPIN-2019-04603) grants; Ontario Centres of Excellence (OCE) Voucher for Innovation and Productivity (VIP1) Program (29119); Fluidigm Canada and the Canada Foundation for Innovation. Alan T. K. Godfrey acknowledges financial support from NSERC's Postgraduate Scholarship - Doctoral and University of Ottawa's Excellence Scholarship.

**Acknowledgements:**

    We acknowledge Dr. Maohui Chen for training in atomic force microscopy and Tony Olivieri for training related to polyimide film fabrication. We are also pleased to acknowledge the support of laboratory engineers Yu-Hsuan Wang and Tyler Clancy from University of Ottawa throughout this work.

**Disclosures:**

The authors declare no conflicts of interest.

# References

[1] M. Lenzner *et al.*, "Femtosecond Optical Breakdown in Dielectrics," *Phys. Rev. Lett.*, vol. 80, no. 18, pp. 4076–4079, May 1998, doi: 10.1103/PhysRevLett.80.4076.

[2] D. M. Rayner, A. Naumov, and P. B. Corkum, "Ultrashort pulse non-linear optical absorption in transparent media," *Opt. Express*, vol. 13, no. 9, pp. 3208–3217, May 2005, doi: 10.1364/OPEX.13.003208.

[3] N. T. Kattamis, P. E. Purnick, R. Weiss, and C. B. Arnold, "Thick film laser induced forward transfer for deposition of thermally and mechanically sensitive materials," *Appl. Phys. Lett.*, vol. 91, no. 17, p. 171120, Oct. 2007, doi: 10.1063/1.2799877.

[4] P. Delaporte and A.-P. Alloncle, "[INVITED] Laser-induced forward transfer: A high resolution additive manufacturing technology," *Opt. Laser Technol.*, vol. 78, pp. 33–41, Apr. 2016, doi: 10.1016/j.optlastec.2015.09.022.

[5] M. S. Brown, C. F. Brasz, Y. Ventikos, and C. B. Arnold, "Impulsively actuated jets from thin liquid films for high-resolution printing applications," *J. Fluid Mech.*, vol. 709, pp. 341–370, Oct. 2012, doi: 10.1017/jfm.2012.337.

[6] N. T. Goodfriend *et al.*, "Blister-based-laser-induced-forward-transfer: a non-contact, dry laser-based transfer method for nanomaterials," *Nanotechnology*, vol. 29, no. 38, p. 385301, Jul. 2018, doi: 10.1088/1361-6528/aaceda.

[7] A. Piqué, H. Kim, and C. B. Arnold, "Laser Forward Transfer of Electronic and Power Generating Materials," in *Laser Ablation and its Applications*, C. Phipps, Ed. Boston, MA: Springer US, 2007, pp. 339–373.

[8] M. S. Brown, N. T. Kattamis, and C. B. Arnold, "Time-resolved study of polyimide absorption layers for blister-actuated laser-induced forward transfer," *J. Appl. Phys.*, vol. 107, no. 8, p. 083103, Apr. 2010, doi: 10.1063/1.3327432.

[9] N. T. Kattamis, N. D. McDaniel, S. Bernhard, and C. B. Arnold, "Laser direct write printing of sensitive and robust light emitting organic molecules," *Appl. Phys. Lett.*, vol. 94, no. 10, p. 103306, Mar. 2009, doi: 10.1063/1.3098375.

[10] D. L. N. Kallepalli *et al.*, "Multiphoton laser-induced confined chemical changes in polymer films," *Opt. Express*, vol. 28, no. 8, pp. 11267–11279, Apr. 2020, doi: 10.1364/OE.389215.

[11] O. H. Y. Zalloum, M. Parrish, A. Terekhov, and W. Hofmeister, "On femtosecond micromachining of HPHT single-crystal diamond with direct laser writing using tight focusing," *Opt. Express*, vol. 18, no. 12, pp. 13122–13135, Jun. 2010, doi: 10.1364/OE.18.013122.

[12] K. Sugioka, "Progress in ultrafast laser processing and future prospects," *Nanophotonics*, vol. 6, no. 2, pp. 393–413, 2016, doi: 10.1515/nanoph-2016-0004.

[13] C. B. Schaffer, A. Brodeur, and E. Mazur, "Laser-induced breakdown and damage in bulk transparent materials induced by tightly focused femtosecond laser pulses," *Meas. Sci. Technol.*, vol. 12, no. 11, pp. 1784–1794, Oct. 2001, doi: 10.1088/0957-0233/12/11/305.

[14] M. Nishikawa, B. Taheri, and J. L. West, "Mechanism of unidirectional liquid-crystal alignment on polyimides with linearly polarized ultraviolet light exposure," *Appl. Phys. Lett.*, vol. 72, no. 19, pp. 2403–2405, May 1998, doi: 10.1063/1.121390.

[15] B. Li, T. He, and M. Ding, "Tuning the aggregation of polyimide thin films by modification of their molecular interactions," *Polym. Int.*, vol. 49, no. 4, pp. 395–398, Apr. 2000, doi: 10.1002/(SICI)1097-0126(200004)49:4<395::AID-PI392>3.0.CO;2-2.

[16] "DuPont Kapton -- Summary of Properties." DuPont, [Online]. Available: https://www.dupont.com/content/dam/dupont/products-and-services/membranes-and-films/polyimde-films/documents/DEC-Kapton-summary-of-properties.pdf.

[17] "PRODUCT BULLETIN - PI 2525, PI 2555 & PI 2574." HD Microsystems, Nov. 2012, [Online]. Available: http://web.mit.edu/scholvin/www/mq753/Documents/resists.PI-2525_2555_2574_ProductBulletin.pdf.

[18] M. G. Vivas, T. Shih, T. Voss, E. Mazur, and C. R. Mendonca, "Nonlinear spectra of ZnO: reverse saturable, two- and three-photon absorption," *Opt. Express*, vol. 18, no. 9, pp. 9628–9633, Apr. 2010, doi: 10.1364/OE.18.009628.

[19] S. H. Glenzer *et al.*, "Observations of Plasmons in Warm Dense Matter," *Phys. Rev. Lett.*, vol. 98, no. 6, p. 065002, Feb. 2007, doi: 10.1103/PhysRevLett.98.065002.

[20] D. L. N. Kallepalli, A. M. Alshehri, D. T. Marquez, L. Andrzejewski, J. C. Scaiano, and R. Bhardwaj, "Ultra-high density optical data storage in common transparent plastics," *Sci. Rep.*, vol. 6, May 2016, doi: 10.1038/srep26163.

[21] K. L. N. Deepak, R. Kuladeep, S. Venugopal Rao, and D. Narayana Rao, "Luminescent microstructures in bulk and thin films of PMMA, PDMS, PVA, and PS fabricated using femtosecond direct writing technique," *Chem. Phys. Lett.*, vol. 503, no. 1, pp. 57–60, Feb. 2011, doi: 10.1016/j.cplett.2010.12.069.

[22] Z. Nie *et al.*, "Multilayered optical bit memory with a high signal-to-noise ratio in fluorescent polymethylmethacrylate," *Appl. Phys. Lett.*, vol. 94, no. 11, p. 111912, Mar. 2009, doi: 10.1063/1.3103365.

[23] A. M. Alshehri, K. L. N. Deepak, D. T. Marquez, S. Desgreniers, and V. R. Bhardwaj, "Localized nanoclusters formation in PDMS upon irradiation with femtosecond laser," *Opt. Mater. Express*, vol. 5, no. 4, pp. 858–869, Apr. 2015, doi: 10.1364/OME.5.000858.

[24] V. Sametoglu, V. T. K. Sauer, and Y. Y. Tsui, "Production of 70-nm Cr dots by laser-induced forward transfer," *Opt. Express*, vol. 21, no. 15, pp. 18525–18531, Jul. 2013, doi: 10.1364/OE.21.018525.

# Appendix A – Derivation & Full Summary of Calculation

This appendix provides the complete details and parameters used for the calculations shown in the main paper. These calculations were performed using Mathematica version 11.1.

The advantages of using femtosecond lasers to induce material breakdown are two-fold. Femtosecond pulses deposit energy with minimal heat dissipation, and because amplified femtosecond pulses are extremely intense, nonlinear absorption processes are easily accessible. These two factors, when combined, lead to deposition of energy with much greater confinement than other methods.

*Table 1 – Summary of Simulated Polymer and Laser Parameters used in Calculation*

| Symbol | Definition | Value | Reference[LNDK1][LNDK2] |
|---|---|---|---|
| $\rho$ | Density of polyimide | $1.42$ g cm$^{-3}$ | [16] |
| $c$ | Specific heat capacity of polyimide | $1.09$ J g$^{-1}$ K$^{-1}$ | [17] |
| $T_{decomp}$ | Decomposition temperature of polyimide | 550°C | [17] |
| $\alpha$ | 1PA absorption coefficient of polyimide at $\lambda = 355$ nm | $1.3 \times 10^4$ cm$^{-1}$ | [3] |
| $\beta$ | 2PA absorption coefficient of ZnO at $\lambda \sim 550 - 750$ nm | $\sim 1$ cm/GW | [18] |
| $\gamma$ | 3PA absorption coefficient of ZnO at $\lambda \sim 850 - 950$ nm | $\sim 10^{-3}$ cm$^3$/GW$^2$ | [18] |
| $\tau$ | Duration of pulse, square envelope | 50 fs | chosen |
| $\omega_0$ | Radius of beam waist | 1.5 µm | chosen |
| $E_{pulse}$ | Pulse energies used in each case (1PA, 2PA, 3PA) | 42, 31, 2.5 nJ | chosen |

    We demonstrate this in the following calculation of femtosecond-laser-induced breakdown through linear, two-photon, and three-photon absorption (henceforth 1PA, 2PA, and 3PA) in a polymer film. Each mechanism was considered separately, as is typical by material selection in LIFT experiments using polymer films, and which naturally highlights the main features of energy deposition from each absorption order. We model the film to have the density, heat capacity and decomposition temperature of polyimide [16], [17]. The linear absorption coefficient was chosen to match experiments involving blister formation through linear absorption of 355 nm light in polyimide [3]. Since nonlinear absorption coefficients of undoped polymers are not reported, we chose two- and three-photon absorption cross sections to be those reported for zinc oxide, a material with a similar band gap

[18]. In all three scenarios, laser pulses were taken to have durations of $\tau = 50$ fs with square envelopes and Gaussian spatial intensity profiles with focal spot diameters of $2\omega_0 = 3$ μm at their waists. The focal spot was placed at the surface of the modelled polymer films in all cases. These values are summarized in **Table 1**.

Since femtosecond pulses are much shorter than timescales of heat dissipation, absorption is approximated to be instantaneous. Thus, pulse intensities incident on the polymer are taken to be $I_0(r) = I_{peak} \cdot e^{-2r^2/\omega_0^2}$ where peak intensity in space and time is defined as $I_{peak} = \frac{2}{\pi \omega_0^2} \frac{E_{pulse}}{\tau}$. Pulse energies were chosen separately for the cases of 1PA, 2PA, and 3PA, such that the peak temperature in the material reached 10000K in all cases; these pulse energy values were 42 nJ, 31 nJ, and 2.5 nJ respectively. This was done to compare the confinement of similar levels of breakdown for each absorption order.

The intensity attenuation from each absorption mechanism was calculated and converted into energy deposition proportional to $-\frac{dI}{dz}$ where $z$ is the direction of propagation into the polymer film. Circular symmetry was used to simplify these calculations. In particular, the energy absorbed within a small annular cylinder of thickness with an inner radius of $r$, outer radius of $r + \Delta r$, and a height of $\Delta z$ at a distance $z$ into the film is given by the equation below.

$$E_{abs}(z, \Delta z, r, \Delta r) = 2\pi\tau \cdot \int_r^{r+\Delta r} r' \cdot [I(z, r') - I(z + \Delta z, r')] dr' \quad (1)$$

The function $I(z, r)$ represents the intensity at each point in the material, given by the below equations in the cases of 1PA, 2PA, and 3PA.

$$I(z, r) = I_0(r) \cdot \begin{cases} e^{-\alpha z} & \text{for 1PA} \\ [(1 + \beta \cdot I_0(r) \cdot z)^{-1}] & \text{for 2PA} \\ [1 + 2\gamma \cdot I_0^2(r) \cdot z]^{-1/2} & \text{for 3PA} \end{cases} \quad (2)$$

To obtain local temperature increase $\Delta T$, we equated the energy absorbed to heat energy as given by $Q = \rho V c \Delta T$ where $V = \pi h \Delta r [2r + \Delta r]$ is the volume of each annular cylindrical region. Therefore, the local temperature increase is given by:

$$\Delta T(z, \Delta z, r, \Delta r) = \frac{1}{\pi} \cdot \frac{E_{abs}(z, \Delta z, r, \Delta r)}{c\rho \Delta z \Delta r [2r + \Delta r]} \quad (3)$$

Using Equation (3), we generated plots of the temperature distributions induced by 1PA, 2PA, and 3PA respectively. These plots are shown in **Fig. 1** of the main paper. Regions where the local temperature exceeds the breakdown temperature are shown by contours. The breakdown region is thinner and shallower in the cases of nonlinear absorption. Most notably, the energy deposition is over 20x shallower in the case of 3PA, compared to 1PA, and approximately 1.5x more confined in width. This illustrates the two advantages of using nonlinear absorption for creating blister microstructures: {1} very shallow penetration depths which allow for more precise energy deposition in a thin film and {2} the possibility to confine energy to areas below the focal spot size due to inherent thresholding effects. The former is important for controlling the expansion process of a blister, and the latter offers the potential for direct fabrication of blister structures below the diffraction limit.

**Appendix B – Blister Rupture at High Pulse Energies**

In blister fabrication experiments, excessive pulse energies result in mechanical failure of the film. This is caused by excessive laser penetration into the film and/or pressure built up beneath the film. AFM data of ruptured blisters in a 1.3 µm polyimide film are shown in **Figure 8**. Between 225 nJ and 375 nJ of pulse energy, blisters were no longer left intact; rupture was seen in the form of infrequent cracking and reduction in blister height. At higher energies, blisters show consistent cracking and in extreme cases, partial removal of the film.

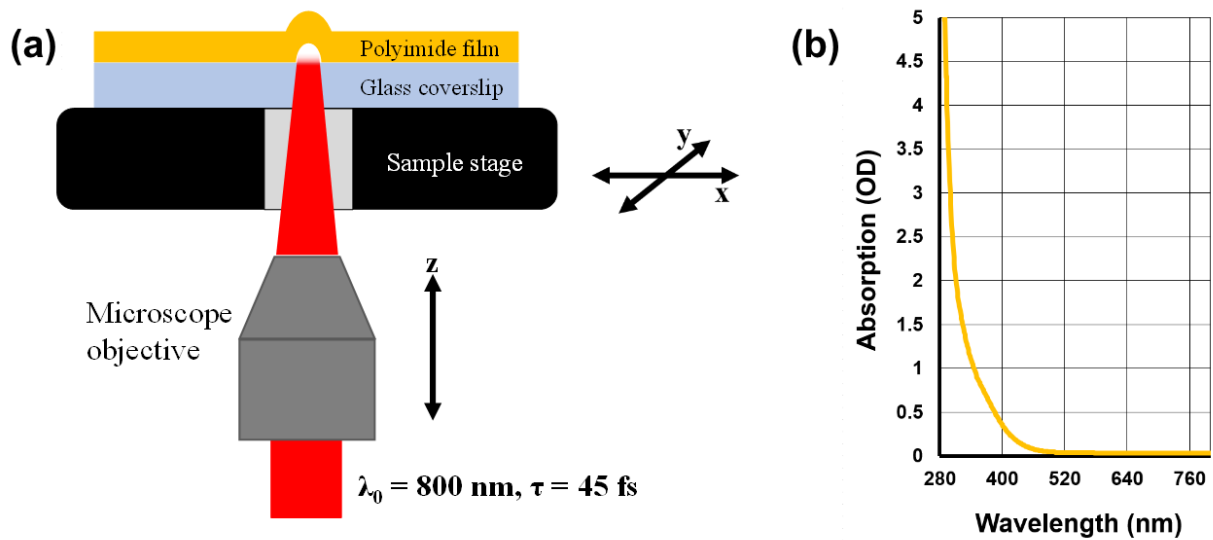

**Figure 1** – **(a)** Schematic of experimental configuration. Single fs pulses are focused through a coverslip substrate, onto the underside of a polyimide film. **(b)** Visible linear absorption spectrum of polyimide film with a thickness of 1.3 µm. No linear absorption was seen for visible wavelengths longer than 500 nm.

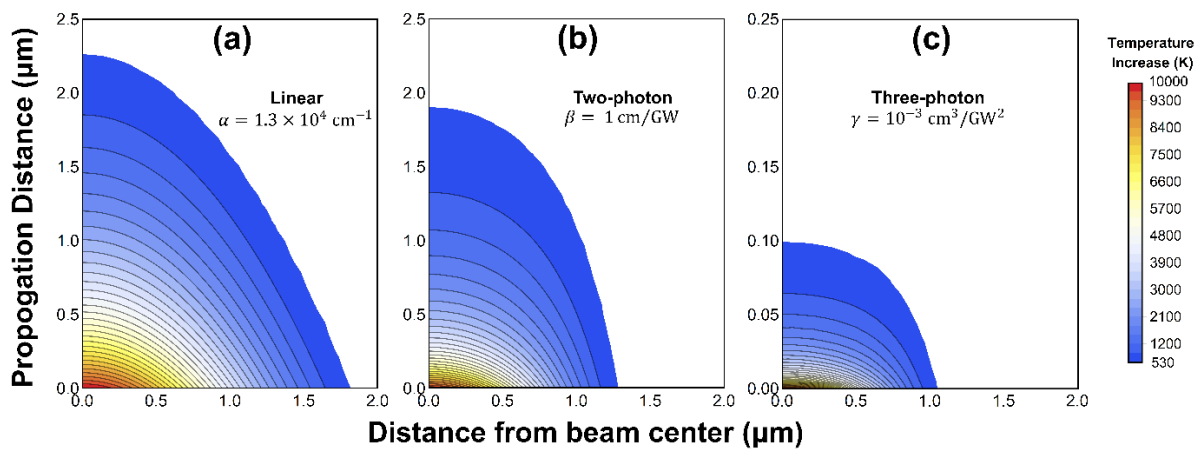

**Figure 2** – Calculated temperature distributions induced by a 50-fs laser pulse focused to a 1.5 µm radius in materials with **(a)** linear (polyimide-like), **(b)** two-photon (ZnO-like), and **(c)** three-photon (ZnO-like) absorption mechanisms. Pulse energies of **(a)** 42 nJ, **(b)** 31 nJ, and **(c)** 2.5 nJ were chosen to set the peak temperature to 10000 K in all cases. The vertical axis for (c) is stretched by a factor of 10.

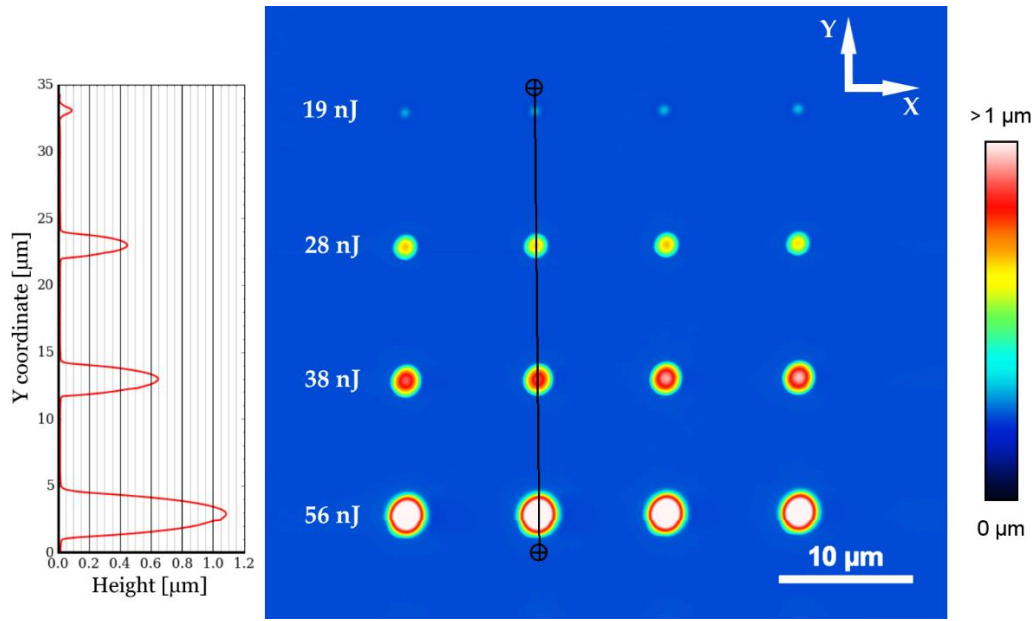

**Figure 3** – An AFM image of blisters fabricated in 1.3 µm polyimide using a 0.4 NA objective, with pulse energies near the damage threshold. The AFM line profile displayed (left) corresponds to the cross-section denoted by the black line in the 2D image. The blister formation threshold was seen at 19 nJ of pulse energy (peak intensity of 4 x $10^{12}$). Near the threshold, small changes in pulse energy create drastic changes in blister height and diameter.

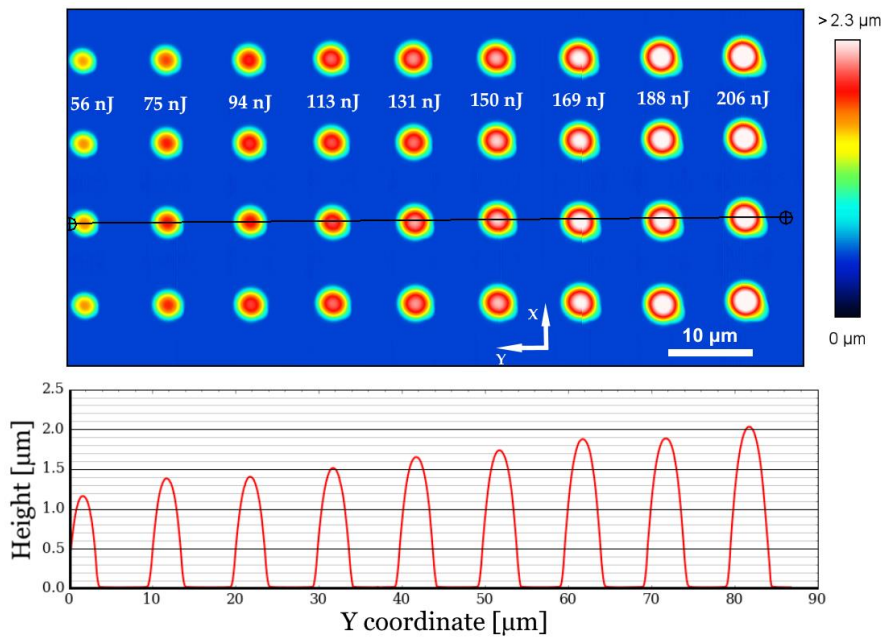

**Figure 4** – An AFM image of blisters fabricated in 1.3 µm polyimide using a 0.4 NA objective with increased pulse energies. The AFM line profile displayed (bottom) corresponds to the cross-section denoted by the black line in the 2D image. For intermediate pulse energies, blister heights and diameters scale less steeply with pulse energy.

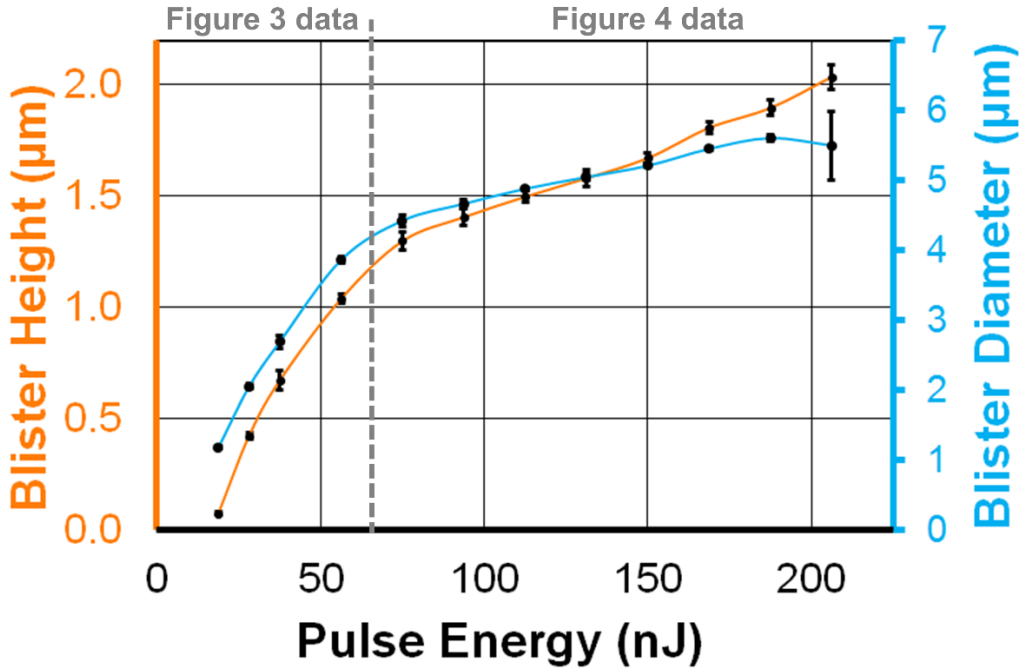

**Figure 5** – Blister height and diameter dependence on pulse energy for blisters. As seen in **Fig 3**, near-threshold scaling of height and diameter is nonlinear. At the higher energies shown in **Fig. 4**, the trend becomes linear until the onset of blister rupture.

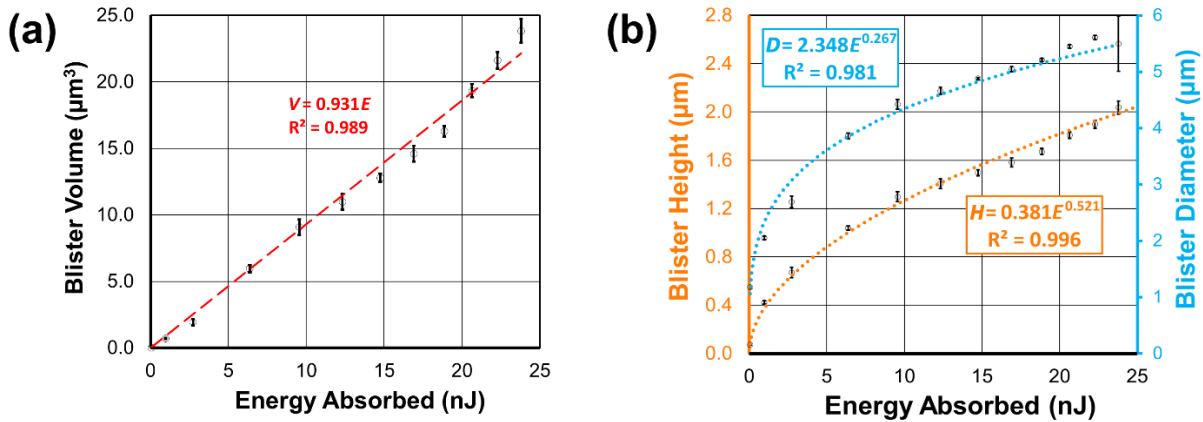

**Figure 6** – **(a)** Blister volume ($V$) and **(b)** blister height ($H$) and diameter ($D$) dependence on pulse energy absorbed in a 1.3 µm polyimide film, using a 0.4 NA objective. The coefficients of determination ($R^2$ ~1) show excellent agreement to the linear fit in **(a)** and power fits in **(b)**.

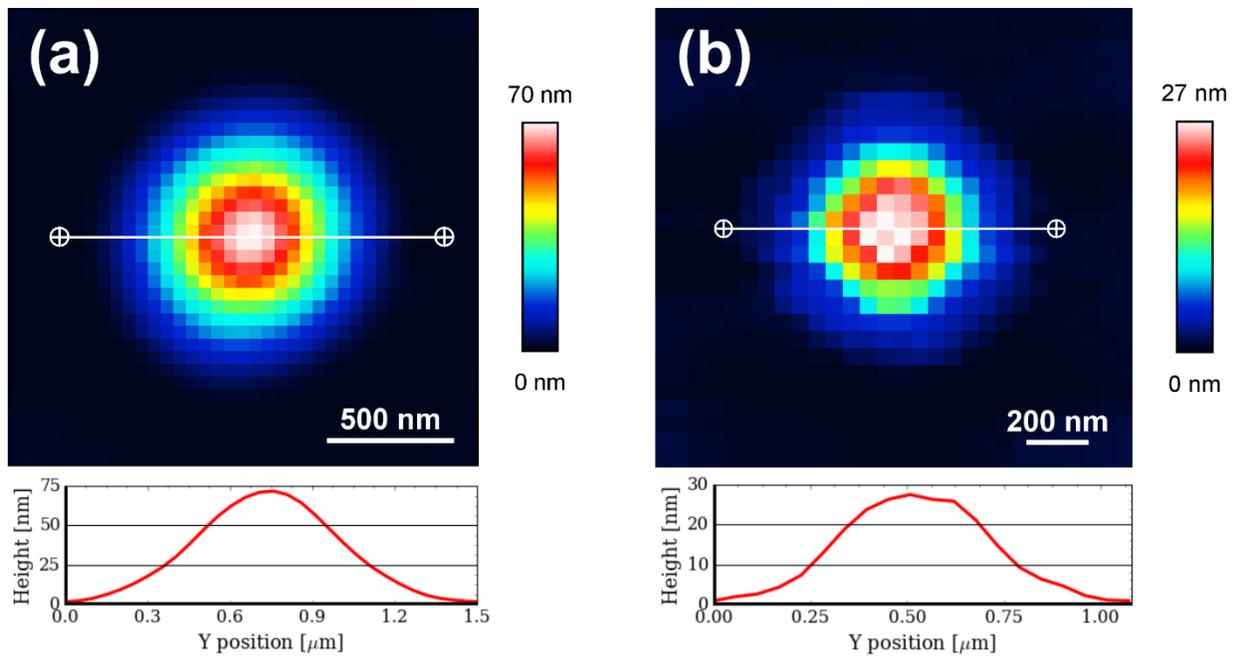

**Figure 7 – (a)** An AFM image and cross-section of the smallest observed blister fabricated using a 0.4 NA objective. The $\frac{1}{e^2}$ diameter of the blister is 1.01 µm, which is 28% of the focal spot diameter. **(b)** An AFM image of the smallest observed blister fabricated using a 0.95 NA objective. The $\frac{1}{e^2}$ diameter is 710 nm, which is 28% of the Gaussian focal spot diameter.

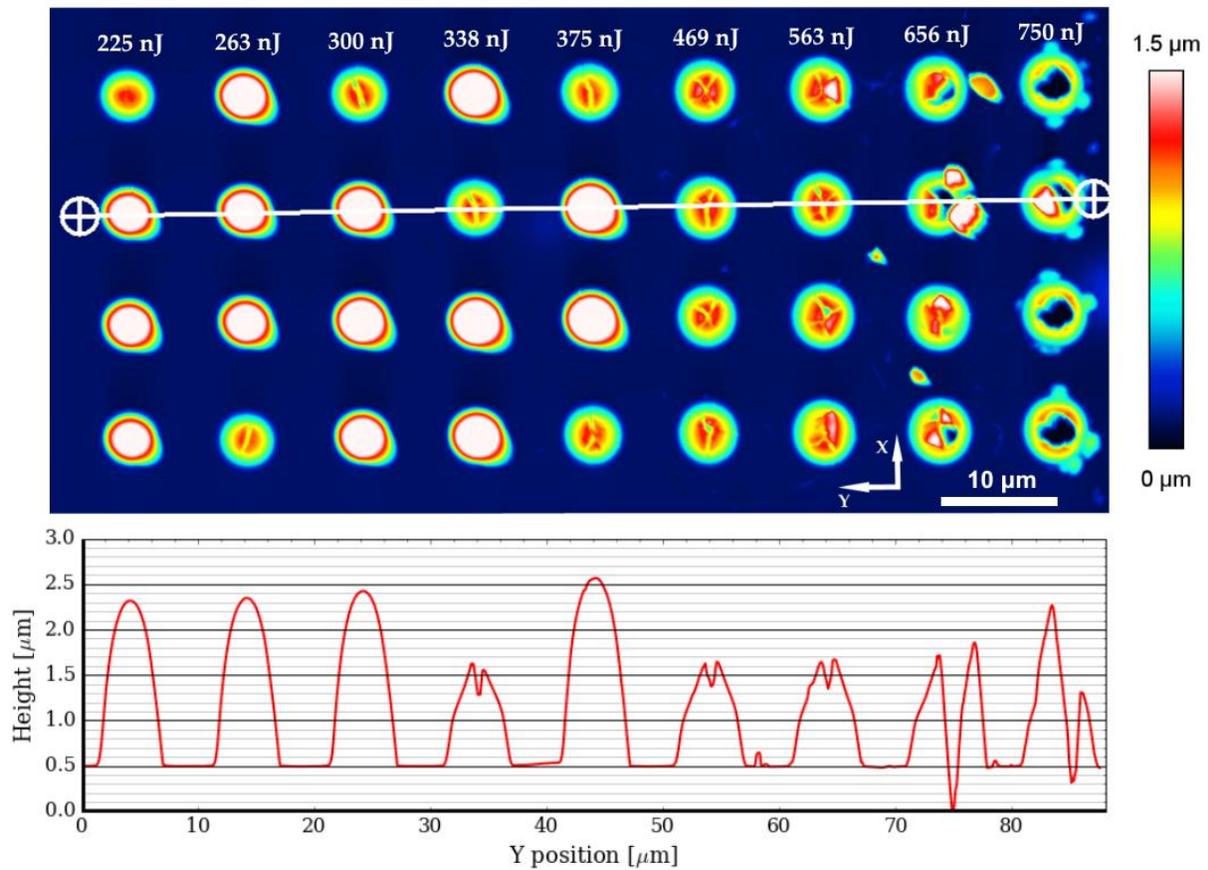

**Figure 8** – An AFM image of blisters fabricated in 1.3 µm polyimide using a 0.4 NA objective with excessive pulse energies. The AFM line profile displayed (bottom) corresponds to the cross-section denoted by the black line in the 2D image. From 225 nJ to 375 nJ, structures show occasional cracking and diminished height. As pulse energy is increased further, blisters show consistent cracking of material, and then material removal.